\documentclass[twocolumn,amsmath,amssymb,prl,showpacs]{revtex4}
\usepackage{setspace}
\usepackage{graphicx}
\usepackage{amsmath}
\usepackage{amssymb}
\usepackage{latexsym}

\usepackage{natbib}
\begin{document}

\title{The dynamics of coherence between coupled and decoupled one-dimensional quasicondensates}
\author{R.G. Scott$^{1,2}$, D.A.W. Hutchinson$^{2}$}
\affiliation{$^{1}$Midlands Ultracold Atom Research Centre, School of Physics and Astronomy, University of Nottingham, Nottingham, NG7 2RD, United Kingdom. \\$^{2}$The Jack Dodd Centre for Quantum Technology, Department of Physics, University of Otago, P.O. Box 56, Dunedin, New Zealand.}
\date{21/05/09}



\begin{abstract}
We reproduce the sub-exponential decoherence of one-dimensional quasicondensates observed in recent experiments. Counter-intuitively, the quasicondensates may decohere even when stongly coupled, if the temperature is large enough or the peak density is low enough to allow significant density fluctuations. We also propose an experiment to investigate the growth of coherence between two initially incoherent quasicondensates. We predict that the coherence will rise on a much slower timescale, and the final coherence again depends strongly on the density fluctuations.
\end{abstract}

\maketitle


Questions of coherence are central to the study of Bose-Einstein condensates (BECs). These questions are interesting from the perspective of fundamental science, but also are crucial for the development of future applications, possibly based on atom chips or optical lattices~\cite{fortaghrev}. Typically, these devices create very elongated clouds, and can realise a quasicondensate regime at temperatures above a characteristic temperature $T_\phi$ that may be much less than the transition temperature $T_c$~\cite{Hofferberth,Hofferberth2,petrov,aspect}. In this regime, global phase coherence is lost, and the phase fluctuates over a characteristic lengthscale which is less than the longitudinal length of the cloud. It remains an open question whether these quasicondensates could be exploited in practical applications, although experiments have attempted to perform interferometry in this regime~\cite{kettnew}. An understanding of the coherence between coupled and decoupled quasicondensates, as well as physical insight into their dynamics, is an important step towards their future use in interferometers. These dynamics are even more intruiging because quasicondensates are often also quantum one-dimensional systems, which sets important restrictions on their behaviour.  

In this letter, we study a recent experiment which realised one-dimensional quasicondensates on an atom chip. The one-dimensional regime occurs when $k_{B}T$ and $U_0 n_p$ are both less than $\hbar \omega_{r}$~\cite{Hofferberth,Hofferberth2}, where $T$ is the temperature of the cloud, $n_p$ is its peak density, $\omega_{r}$ is the radial (high) trapping frequency, and $U_0 = 4 \pi \hbar a/m$, in which $a$ is the s-wave scattering length and $m$ is the mass of a single atom in the cloud. For the parameters in the experiment, $U_0 n_p = 2 \times 10^{-30} < \hbar \omega_{r} = 3 \times 10^{-30}$ J. The highest temperature considered in this paper is 200 nK, at which $k_{B}T = 3 \times 10^{-30}$ J. However, when considering the dynamics of splitting, merging and weak links, it is important to question what we mean by one-dimensional. Clearly, during a splitting process into two separate one-dimensional clouds, the originally one-dimensional quasicondensate passes through a two-dimensional intermediate stage. We address this ambiguity by performing simulations in which the one-dimensional nature of the quasicondensate is \textit{not} assumed. We find that this two-dimensional state provides a route for equilibration, which may be suppressed in one-dimensional clouds~\cite{weiss}. In fact, counter-intuitively, two quasicondensates may decohere despite being strongly coupled, if the clouds contain sufficiently large density fluctuations. This effect may prevent one from attaining the adiabatic limit when merging quasicondensates in a double well or optical lattice~\cite{polkovnikov,polkov}. This effect also implies that the details of splitting and merging protocols may have a large impact on the final state of the cloud. For example, a splitting or merging protocol which involves a long-lived coupled state may cause significant decoherence between the two quasicondensates, which could be observed as heating due to vortex production~\cite{scottjuddinterf,scottjuddinterf2}. Once a dynamical equilibrium between the two coupled quasicondensates has been reached, we may then recreate the decoherence dynamics~\cite{burkov,mazets,Bistritzer} in the experiment by abruptly severing the link. In this way, we obtain the sub-exponential decay observed experimentally~\cite{Hofferberth} and predicted theoretically~\cite{burkov,mazets}. We also propose an experiment to investigate the growth of coherence between two initially incoherent quasicondensates when they are merged to create a link. At low temperatures, the coherence rises, but on a slow timescale, which we explain in terms of vortex dynamics, in an analagous effect to the Kibble-Zurek mechanism~\cite{PolkovnikovKZ}. At higher temperatures, the presence of density fluctuations inhibits the growth of coherence.

The results in this paper are obtained using the finite-temperature truncated Wigner method, which we developed in a previous publication to analyse atom-chip interferometry of BECs~\cite{scottjuddinterf2}. This method has the advantage of capturing the zero-temperature mean-field depletion, which has a significant effect on high density elongated clouds. Essentially, the method models thermal fluctuations by calculating the Bogoliubov excitations of the condensate mode, and then populating them with a thermal distribution (see Ref.~\cite{scottjuddinterf2} for details). As in our previous paper~\cite{scottjuddinterf2}, we do not add excitations with an average occupation of less than one atom. The exclusion of vacuum fluctuations and hence incoherent scattering allows us to clearly identify the physical mechanisms of decoherence in our calculations, and also it eliminates any spurious decoherence caused by the numerical simulation of quantum noise in the truncated Wigner method~\cite{foot1}. To match with the experiments, we construct finite $T$ initial states containing $N_{T} = 5\times 10^{3}$ $^{87}$Rb atoms, and the trap frequencies in the axial and radial directions are $\omega_{z} = 2\pi \times 5$ rad s$^{-1}$ and $\omega_{r} = \omega_{x} = \omega_{y} = 2\pi \times 4000$ rad s$^{-1}$ respectively. Once we have prepared three-dimensional initial states, we evolve in time a two-dimensional slice in the $y=0$ plane, since the motion in the $y$-direction is frozen out.

\begin{figure}[tbp]
\centering
\includegraphics[width=1.0\columnwidth]{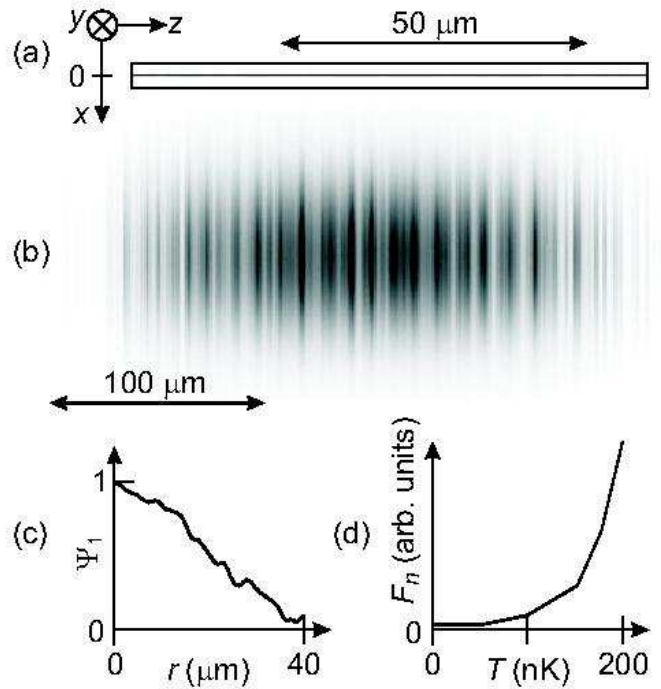}
\caption{(a) Typical atom density profile of central region ($\sim 1/3$ of total cloud) of quasicondensate in $y=0$ plane (axes inset) at 0 nK. (b) Corresponding atom density profile of complete cloud after 8 ms expansion. Upper [lower] horizontal bar indicates scale in (a) [(b)]. (c) Coherence $\Psi_{1}$ as a function of distance $r$ in the $z$-direction within the quasicondensate at 0 nK. (d) In-trap density fluctuations $F_{n}$ as a function of $T$.}
\label{f1}
\end{figure}

We begin by characterising a single cloud at 0 K. Figure~\ref{f1}(a) shows the density profile in the $y=0$ plane of the central section (approximately one third) of the cloud. The figure illustrates the extreme aspect ratio of the system. Such clouds are commonly imaged after expansion, during which they expand rapidly in the radial direction. Figure.~\ref{f1}(b) shows the full cloud after 8 ms expansion. For the parameters in the experiment, $T_{\phi} = 15N_{T}\left(\hbar\omega_{z}\right)^{2}/32\mu k_{B} \approx 0.5$ nK~\cite{petrov,aspect}, where $\mu = 3.5 \times 10^{-30}$ J is the chemical potential. This temperature is much less than $T_c = 330$ nK. Consequently, we would expect to find a true BEC with global phase coherence at 0 nK. To test this, we calculate the coherence of the cloud $\Psi_{1}(r)$, where $r$ is distance in the $z$-direction from some arbitrary point $(0,0,z)$ [axes inset in Fig.~\ref{f1}(a)] near the center of the cloud. This is done by comparing the phase $\theta(r+r_n)$ to $\theta(r_n)$, where $r_n$ are $N$ positions near the center of the cloud. We may define $\Psi_{1}(r)$ mathematically as
\begin{equation}
\Psi_{1}(r) = \frac{1}{N} \left| \sum^{N}_{n=1} e^{i\left\{\theta(r+r_n)-\theta(r_n)\right\}} \right| .
\label{coherence}
\end{equation}
The resulting plot of $\Psi_{1}(r)$ is shown in Fig.~\ref{f1}(c). The curve shows that coherence drops from one to almost zero over $\sim40$ $\mu$m. This is much less than the longitudinal length of the cloud, which we calculate to be $\sim250$ $\mu$m. We conclude that even the zero-$T$ mean-field depletion is sufficient to destroy global phase coherence in this extremely dense and elongated cloud.

Although quasicondensates have no global phase coherence, density fluctuations are suppressed due to the mean-field interaction~\cite{petrov,aspect,schumm}. For example, the quasicondensate shown in Fig.~\ref{f1}(a) has a reasonably smooth density profile. Density fluctuations may be observed following expansion, when interactions become negligible, as shown in Fig.~\ref{f1}(b). However, as the temperature is increased, small density fluctuations may appear in-trap. We may characterise these density fluctuations with the quantity $F_{n}\left(T\right)$, defined as
\begin{equation}
F_{n}\left(T\right) = \frac{ \frac{1}{2L}\int^{L}_{-L} n^{2}\left(z,T\right) dz - 
\left[ \frac{1}{2L}\int^{L}_{-L} n\left(z,T\right) dz \right]^{2} }
{\left[ \frac{1}{2L}\int^{L}_{-L} n\left(z,T\right) dz \right]^{2}},
\label{denflucs}
\end{equation}
where $n\left(z,T\right)$ is atom density at position $(0,0,z)$ and temperature $T$, and $2L$ is a distance less than the axial length of the cloud. We plot $F_{n}\left(T\right)$ in Fig.~\ref{f1}(c) for $2L=20$ $\mu$m. The curve shows that density fluctuations are suppressed at low temperatures, but grow rapidly when the temperature is raised above 100 nK.

\begin{figure}[tbp]
\centering
\includegraphics[width=1.0\columnwidth]{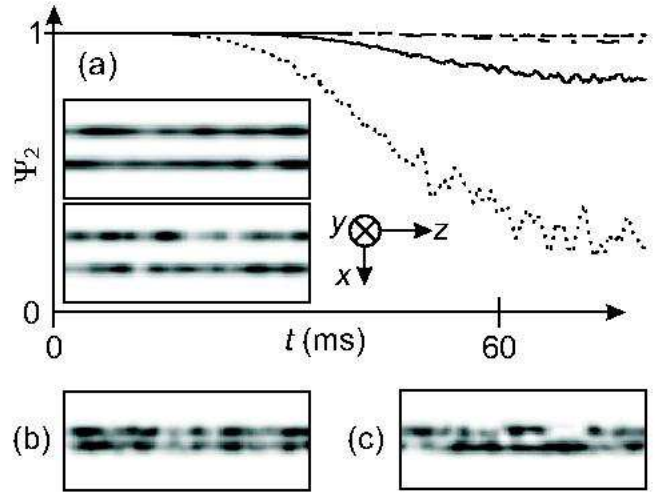}
\caption{(a) Coherence $\Psi_{2}$ plotted as a function of time for two coupled/uncoupled quasicondensates at 100 nK (solid curve/dashed curve) and 200 nK (dotted curve/dot-dashed curve). Upper (lower) inset shows the atom density profile in the central region of two coupled clouds in the $y=0$ plane (axes inset) at $t=50$ ms and $T=100$ (200) nk. The field-of-view is $7.5 \times 3$ $\mu$m. (b) Atom density profile in central region of a quasicondensate at 200 nK at $t=60$ ms, after performing merging protocol A. The field-of-view is identical to that in (a) insets. (c) As (b), but for merging protocol B.}
\label{f2}
\end{figure}

 
We now simulate the dynamic equilibrium of two coupled quasicondensates~\cite{whitlock}, separated by a distance $D$, as a function of temperature. The coherence $\Psi_{2}(t)$ between the two clouds as function of time $t$ is measured as
\begin{equation}
\Psi_{2}(t) = \frac{1}{2L} \left| \int^{L}_{-L} e^{i \vartheta \left(z\right)} dz \right| ,
\label{coherence2}
\end{equation}
where $\vartheta \left(z\right)$ is the relative phase between the two clouds, and $2L$ is 100 $\mu$m. This quantity is always averaged over five simulations with different initial conditions. We begin our simulations with two identical quasicondensates, but with a very small amount of added noise in order to provide a seed for the dynamic evolution. This noise is on average much less than half a particle per mode, which was added in previous work to model incoherent scattering~\cite{norrieprl,meotago,meCheren}. Consequently, incoherent scattering is prohibited in our calculations. We check this condition is satisfied by calculating $\Psi_{2}(t)$ for $D= 1.5$ $\mu$m, which is sufficiently large to suppress coupling between the two clouds. In this case, the coherence does not drop significantly from one over 80 ms, as shown in Fig.~\ref{f2}(a) for 100 nk (dashed curve) and 200 nk (dot-dashed curve). Of course, in a real experiment incoherent scattering \emph{would} cause decoherence, but our purpose here is to prepare coupled quasicondensates in dynamical equilibrium. To do this, we set $D$ to 1 $\mu$m, so that the height of the barrier is roughly equal to the chemical potentials of the clouds. Counter-intuitively, the coupling \emph{increases} the rate of decoherence. This effect is shown by the solid line in Fig.~\ref{f2}(a) at $T=100$ nK. At this temperature, the effect is relatively small, and the coherence plateaus at a value of about 0.8. However, the effect is much more dramatic at 200 nK, where the coherence drops to about 0.2 [dotted curve in Fig.~\ref{f2}(a)]. It is intruiging that this coupled system allows a route to equilibration, which may be suppressed for one-dimensional clouds~\cite{weiss}. These results show that, suprisingly, there is no coherence for coupled quasicondensates above a certain temperature, and hence the experiments would have been impossible at higher temperature, despite being well below $T_c$. 

The decoherence process occurs via small initial differences in phase or density between the two clouds driving small Josephson oscillations, causing a transfer of atoms between the two wells. As atoms are gained or lost, the mean-field interaction in each quasicondensate responds by modifying the density profile. This leads to greater differences in phase or density between the two clouds, driving greater Josephson oscillations. However, this process does not ultimately lead to a complete loss of coherence. Since the Josephson oscillations are incoherent over the phase coherence length of the quasicondensate, they create fluctuations in density over this distance. Ultimately, the amplitude of the Josephson oscillations is restricted by the amplitude of the density fluctuations at the temperature of the cloud. At 100 nK, the density fluctuations are small, as shown in the upper inset of Fig.~\ref{f2}(a), which is an enlargement of the central region of the two coupled clouds at $t=50$ ms. These small fluctuations should be compared to the much larger fluctuations in the corresponding image at 200 nK in the lower inset of Fig.~\ref{f2}(a). As previously shown in Fig.~\ref{f1}(c), the magnitude of the density fluctuations increases rapidly as the temperature is raised above 100 nK.




Our calculations reveal the suprising result that significant decoherence may occur while the quasicondensates are strongly coupled, and consequently the details of splitting and merging protocols may have a large impact on the cloud dynamics. To illustrate this, we consider two protocols in which two initially identical quasicondensates are merged from $D=1.0$ to $D=0.5$ $\mu$m. In protocol A, the clouds are merged over 1 ms and then held for a further 59 ms. In this case, there is little excitation of the cloud, as shown in Fig.~\ref{f2}(b), which is an enlargement of the central region of the cloud at $t=60$ ms. In protocol B, the quasicondensates are held at $D=1.0$ $\mu$m for 59 ms, allowing them to reach dynamical equlibrium, before being merged to $D=0.5$ $\mu$m. In this case the density profile at $t=60$ ms has large density fluctuations, and the enlargement of the central region of the cloud in Fig.~\ref{f2}(c) contains two vortices (white circles in center of figure). This figure shows that vortex production, as observed previously in atom interferometers~\cite{kettnew,scottjuddinterf,scottjuddinterf2}, can contribute to heating of the clouds, despite them being individually one-dimensional. 

\begin{figure}[tbp]
\centering
\includegraphics[width=1.0\columnwidth]{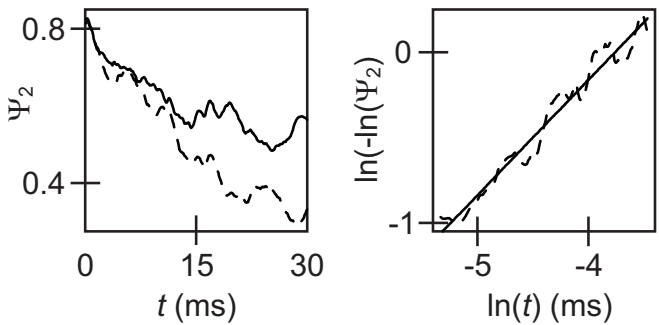}
\caption{(a) Decoherence of two quasicondensates at 100 nK after splitting to $D=1.5$ (dashed curve) and 1.15 (solid curve) $\mu$m. (b) Dashed curve: double logarithmic plot of dashed curve in (a) showing sub-exponential decay of coherence. Solid line: line of best fit.}
\label{f3}
\end{figure}

Figure \ref{f2}(a) shows that our simulations of coupled quasicondensates reach a dynamical equilibrium after 80 ms evolution. For $t\gtrsim80$ ms, our simulations capture the correct variations in the occupations of the Bogoliubov modes of two coupled quasicondensates. We may then simulate the decoherence of the two quasicondensates by abruptly splitting them to $D=1.5$ $\mu$m and consequently severing the link between them, allowing the Bogoliubov modes in each cloud to evolve freely. Hence we avoid the challenging problem of directly modelling the uncertainties in the splitting process in the experiment~\cite{Hofferberth}. The resulting loss of coherence at 100 nK is shown by the dashed curve in Fig.~\ref{f3}, which has a very similar timescale and form to that reported in experiment. We also carry out a partial splitting to $D=1.15$ $\mu$m, and observe the coherence saturate at $\sim 0.6$ (solid curve).

Theory has predicted that the coherence should decay as $e^{\left(-t/t_0\right)^{\alpha}}$, where $\alpha = 2/3$~\cite{burkov,mazets}. We investigate this by plotting $\ln\left(-\ln\left(\Psi_{2}\right)\right)$ against $\ln\left(t\right)$, as shown in Fig.~\ref{f3}(b). The gradient of the line of best fit (solid line) is 0.7, in rough agreement with theory and experiment. We also extract a value of 23 ms for $t_0$ from the same graph, in good agreement with the value predicted by the analytic results in Ref.~\cite{mazets}.



\begin{figure}[tbp]
\centering
\includegraphics[width=0.5\columnwidth]{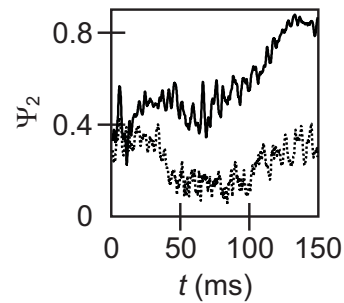}
\caption{Establishment of coherence when two initially incoherent quasicondensates are coupled at 100 nk (solid curve) and 200 nk (dotted curve).}
\label{f5}
\end{figure}

We now extend our analysis to coupling of two initially incoherent quasicondensates. We find that when the two quasicondensates are held at $D=1.0$ $\mu$m at 100 nk, the coherence rises over about 150 ms (solid curve in Fig.~\ref{f5}). Crucially, the growth of coherence when two quasicondensates are coupled occurs on a wildly different timescale to the loss of coherence when two quasicondensates are decoupled (shown in Fig.~\ref{f3}). This can be understood in terms of the underlying physical processes. When two quasicondensates are decoupled, the unequal splitting of the thermally occupied Bogoliubov modes causes decoherence on a relatively fast timescale. When two quasicondensates are coupled, vortices form in low density region between the two clouds at positions where the relative phase happens to be $\pi$~\cite{scottjuddinterf,scottjuddinterf2}. These vortices are pinned by the barrier potential, so coherence may only occur once they have either migrated in the $z$-direction to the edges of the cloud, or collided with vortices of opposite rotation and annihilated. This is relatively slow process, occuring over a timescale of $\sim 150$ ms for the parameters of the experiment. This effect is analgous to the Kibble-Zurek mechanism~\cite{PolkovnikovKZ}, in which vortices may survive for long timescales after a rapid quench across a phase transition. In contrast to the results at 100 nK, at 200 nK the coherence remains low (dotted curve in Fig.~\ref{f5}). This is again due to the density fluctuations which occur above 100 nK.

In summary, we have examined the coherence of coupled and decoupled quasicondensates, characterised the time-dependent dynamics and identified the underlying physical processes. It is important to note that, although we focus on the coherence of two quasicondensates in a double well, our findings can also be applied to splitting and merging in optical lattices. Crucially, our results show that there will be heating when quasicondensates are merged in a two-dimensional optical lattice, preventing one from attaining the adiabatic limit~\cite{polkovnikov,polkov}, if the temperature is high enough or the density low enough to allow significant density fluctuations.






\bibliography{biblio}

\begin{thebibliography}{21}
\bibitem[{foo()}]{foot1}
\bibinfo{note}{The truncated Wigner method models quantum fluctuations by
  adding on average half a particle of noise to each mode of the coherent
  initial state. In a highly elongated cloud, this added noise may create
  significant phase fluctuations causing a spurious decoherence.}
\end{thebibliography}

\end{document}